\documentclass[final,3p,times,twocolumn]{elsarticle}

\usepackage{lineno,hyperref}
\usepackage{graphicx}
\usepackage{booktabs}
\usepackage{multirow}
\usepackage{amsmath}
\usepackage{amsfonts}
\usepackage{natbib}
\usepackage{array}
\modulolinenumbers[5]

\journal{Computer Science Review}

\begin{document}

\begin{frontmatter}

\title{Prediction Methods and Applications in the Science of Science: A Survey}

\author[mymainaddress]{Jie Hou}
\author[mymainaddress]{Hanxiao Pan}
\author[mymainaddress]{Teng Guo}

\author[mysecondaryaddress]{Ivan Lee}
\cortext[mycorrespondingauthor]{Corresponding author}
\ead{ivan.lee@unisa.edu.au}

\author[mymainaddress]{Xiangjie Kong}
\author[mymainaddress]{Feng Xia}

\address[mymainaddress]{Key Laboratory for Ubiquitous Network and Service Software of Liaoning Province, School of Software, Dalian University of Technology, China}
\address[mysecondaryaddress]{School of ITMS, University of South Australia, Mawson Lakes, SA 5095, Australia}

\begin{abstract}
  Science of science has become a popular topic that attracts great attentions from the research community. The development of data analytics technologies and the readily available scholarly data enable the exploration of data-driven prediction, which plays a pivotal role in finding the trend of scientific impact. In this paper, we analyse methods and applications in data-driven prediction in the science of science, and discuss their significance. First, we introduce the background and review the current state of the science of science. Second, we review data-driven prediction based on paper citation count, and investigate research issues in this area. Then, we discuss methods to predict scholar impact, and we analyse different approaches to promote the scholarly collaboration in the collaboration network. This paper also discusses open issues and existing challenges, and suggests potential research directions.
\end{abstract}


\end{frontmatter}

%\linenumbers

\section{Introduction}
\label{sec:introduction}

Following the first introduction of the study in science of science \citep{Bernal1939The}, the topic has evolved as an independent discipline to study the development and social function of science and scientific activities. One of the important research topics in this discipline is to predict the trend of scientific development. Xia et al. \cite{Xia2017Big} introduced four popular free-access scholarly data sets, including AMiner, American Physical Society (APS), DataBase systems and Logic Programming (DBLP) and Microsoft Academic Graph (MAG). Many useful attributes such as papers and their authors, the citation relation among papers, paper's publication years are available in all of these datasets. Many studies of paper citation distribution and paper impact prediction, scholar impact quantification and prediction, evolution of author collaboration networks are carried out with these datasets. The development of big scholarly dataset gives more chances to improve the prediction task in the science of science. Therefore, the field of data-driven study of prediction methods becomes an important topic attracting great attentions from the research community \citep{Clauset2017Data}.

Scholarly data are often used to evaluate scholars' outputs. Many academic evaluation indices have been introduced for assessing different scholarly impacts, such as H-index \citep{hirsch2005index} for scholars, journal impact factor (JIF) \citep{Garfield1972Citation} for journals, and citation count for papers. Apart from identifying scholars' present impact, predicting their future impact is more crucial. An accurate prediction can help young researchers to identify targets for future research directions. To perform prediction in academia, metrics for evaluating the impact of paper, author or venue (where scholars publish their papers, such as journals or conferences) are often used. Although the evaluation methods of the academic achievements have been extensively studied, there are specific phenomena remaining challenging researchers, including citation count imbalance across different disciplines \citep{bornmann2017skewness}, self-citation behaviour in the citation network \citep{Ioannidis2015A, cai2019scholarly}, patterns of research collaboration \citep{wang2017scientific}. These phenomena also bring difficulties to predicting scholar's scientific impact and collaboration because their domain specificity and subjectivity make the underlying patterns difficult to model.

Many researchers tried to solve these problems via different methods, including classical machine learning algorithms, network models and problem specific methods.
In this paper, we present a survey of the emerging field of prediction methods and applications in the science of science to provide a comprehensive review of relevant methods and applications.
% Because of the high relevance between academic papers and scholars, the prediction of paper and scholar's impact can be carried out with same datasets and similar prediction framework. In addition, the cooperation of scholars influences the impact of themselves and their outputs.
In recent years, many researchers have focused on the prediction of scholar and paper's impact. The study of author cooperation network also get popular with the development of network science \cite{kong2019academic}. There is a close relationship between scholars and papers. For example, scholar¡¯s impact is always quantified by the impact of their papers, resulting in some indices such as H-index and author impact factor. In addition, the cooperation of scholars influences the impact of themselves \cite{a17} and their outputs \cite{hurley2013deconstructing}.
So, we summarize the research issues from three perspectives: paper impact prediction, scholar impact prediction and author collaboration prediction. In the paper impact prediction section, we review methods for predicting the paper's citation count and citation relationship. Popular models are briefly introduced, including the models based on stochastic process and statistical learning, along with the methods of link prediction. In the scholar impact prediction section, we discuss some classical metrics to evaluate author's impact at first, followed by a brief discussion of the predictability of these metrics. Then, methods aimed at predicting author's impact measured by these metrics are discussed. In the author collaboration prediction section, we focus on the method of link prediction. Some common types of link prediction algorithm are introduced. Besides, other applications of author collaboration networks are discussed in this section. We aim to provide a specialized understanding of the prediction methods and applications in the science of science, and explore the opportunities and challenges in this field.

The rest of the paper is organized as follows. Section \ref{sec:2} summarizes state-of-the-art models for paper citation prediction. Section \ref{sec:3} introduces methods of scholar impact prediction. Section \ref{sec:4} reviews methods and applications of scholar co-authorship prediction. Section \ref{sec:5} presents current and future challenges that are worth paying attention to. Finally, Section \ref{sec:6} concludes this review.

\section{Paper Impact Prediction}
\label{sec:2}
Citation is often used to evaluate the impact of research papers. Predicting papers' citation count is significant for researchers, especially for young scientists, which is helpful to discovery high quality research papers. The prediction task faces several challenges. Firstly, the pattern of paper citation accumulation is complex. Papers can have different citation distributions even if their citation counts are identical. Secondly, most papers only have a few citations in a short term after being published.
Besides, there are many variables influencing paper's citation counts, making it more difficult to model and to predict the citation distribution.

\subsection{The Machine Learning Approach}
\label{sec:2.1}
Machine learning has attracted growing popularity in data mining and prediction. The normal form for solving the prediction problems in the science of science is to extract features of papers and then put them into a machine learning model to get the final prediction result. It is believed that a paper's quality can be described by features such as authors' impact, number of authors, journal's impact, paper length, as well as citation counts. Besides of these, researchers are still looking for more representative assessment methods to quantify paper's quality.

\subsubsection{Models and Features}
\begin{figure}[htbp]
  \begin{center}
    \includegraphics[width=1.0\linewidth]{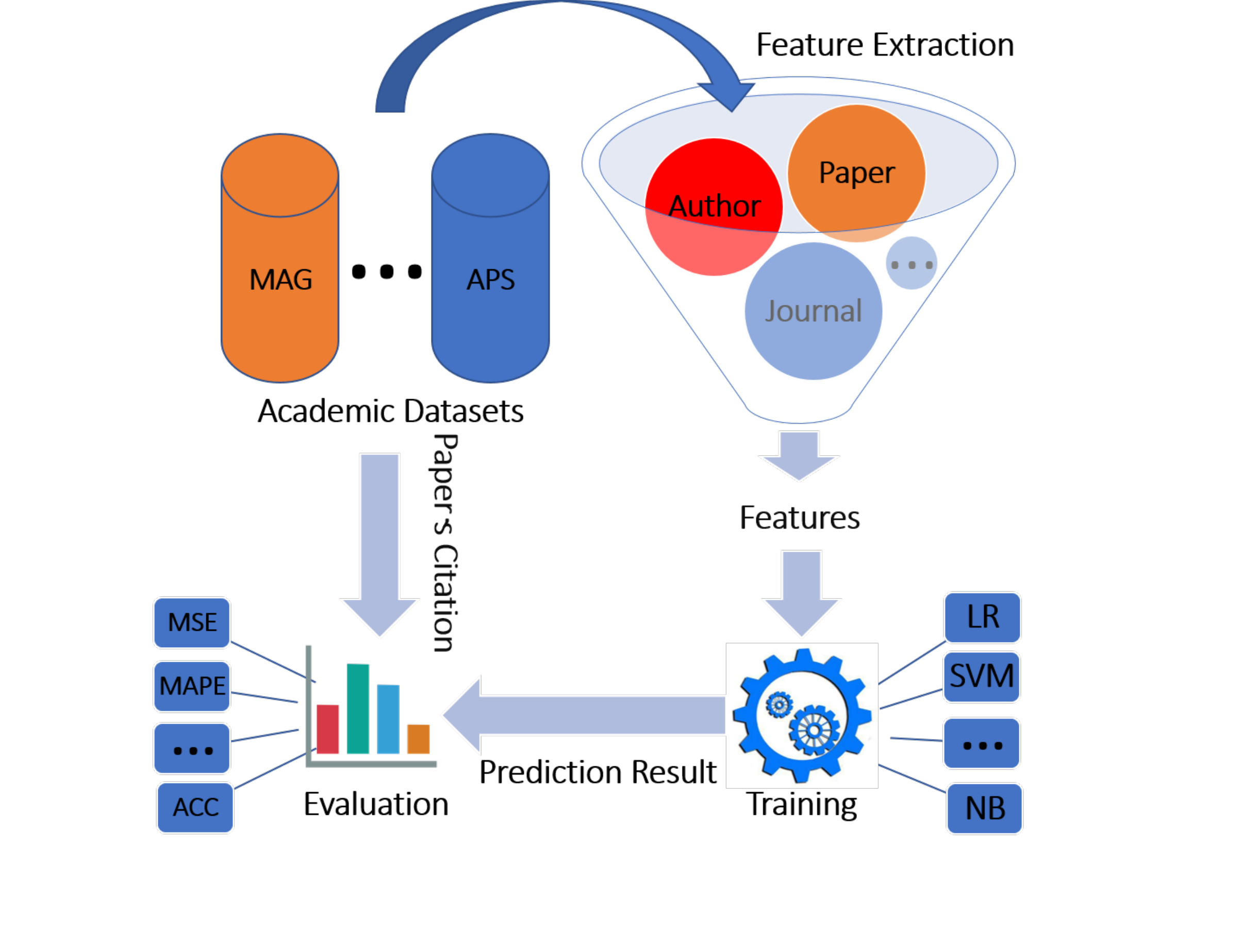}
    \caption{The process of training a predictive model. First, we extract features and labels from the meta data, e.g. the APS dataset, MAG or other scholarly datasets. Then we train a selected machine learning model, such as SVM, NB, LR, and evaluate it with some metrics. Finally, we choose the model performing best to carry out the prediction task.}
    \label{learning}
  \end{center}
\end{figure}

Many researchers have conducted extensive experiments using machine learning models and yielded many achievements \citep{Kong2018VOPRec, Singh2017Understanding, Singh2015The}. Some classical models, e.g. support vector machine (SVM), linear regression (LR), naive Bayessian (NB), and classification and regression tree (CART), are more popular in these works. The overall framework of prediction is described in Fig.\ref{learning}. Compared with other complex models such as deep learning, these models are capable to yield a desirable accuracy in a shorter time.

\newcommand{\tabincell}[2]{
    \begin{tabular}[h]{@{}#1@{}}#2\end{tabular}
}
\begin{table*}
\renewcommand\arraystretch{1.5}
    \centering
    \caption{The Most Commonly Used Features in Paper Citation Prediction}
    \begin{tabular}{c|c|l}
        \hline
        Type & Categories & Features \\
        \hline

        \multirow{10}*{Author-wise} & \multirow{2}*{Citation-based features} & \multirow{2}*{\tabincell{l}{AIF, H-index, g-index, Q-value, author's total\\ citation counts}} \\
        ~ & ~ & ~ \\
        \cline{2-3}
        ~ & \multirow{4}*{Network-based features} & \multirow{4}*{\tabincell{l}{centrality of authors (including degree centrality\\ (DC), betweenness centrality (BC), closeness \\centrality (CC) and eigenvector centrality (EC)),\\ author rank by PageRank or HITS, s-index}} \\
        ~ & ~ & ~ \\
        ~ & ~ & ~ \\
        ~ & ~ & ~ \\
        \cline{2-3}
        ~ &    & \multirow{3}*{\tabincell{l}{academic age, team size, number of co-authors,\\ producibility, institutions attributes (including\\ number of institutions, location, institution rank)}} \\
        ~ & Intrinsic features & ~ \\
        ~ &    & ~ \\
        \cline{2-3}
        ~ & Other features & \multirow{1}*{\tabincell{l}{diversity and distribution of author's topic}} \\
        \hline

        \multirow{6}*{Venue-wise} & \multirow{2}*{Citation-based features} & \multirow{2}*{\tabincell{l}{IF-\emph{n}, total citation counts, H-index, immediacy\\ index, cited half-life, article influence score}} \\
        ~ &    & ~ \\
        \cline{2-3}
        ~ & \multirow{2}*{Network-based features} & \multirow{2}*{\tabincell{l}{centrality of venue (BC, CC, DC, and EC),venue\\ rank by PageRank or HITS, eigenfactor score}} \\
        ~ & ~ & ~ \\
        \cline{2-3}
        ~ & Intrinsic features & \multirow{1}*{\tabincell{l}{number of authors and publications of venue}} \\
        \cline{2-3}
        ~ & Other features & \multirow{1}*{\tabincell{l}{topic diversity and distribution of venue}} \\
        \hline

        \multirow{7}*{Paper-wise} & \multirow{3}*{Citation-based features} & \multirow{3}*{\tabincell{l}{citation of first $n$ years, numbers of authors,\\ countries, institutions or journals/conferences \\citing this paper}} \\
        ~ & ~ & ~ \\
        ~ & ~ & ~ \\
        \cline{2-3}
        ~ & Network-based features & \multirow{1}*{\tabincell{l}{paper rank by PageRank or HITS}} \\
        \cline{2-3}
        ~ &    & \multirow{3}*{\tabincell{l}{years since publication, number of references, \\topic diversity and distribution of paper, topic \\popularity of paper, citation context, paper length}} \\
        ~ & Intrinsic features & ~ \\
        ~ &    & ~ \\
        \hline
    \end{tabular}
\label{features}
\end{table*}

Commonly used features in the citation prediction can be divided into three categories: paper-wise features, author-wise features and venue-wise features. The details are shown in Table \ref{features}. Since most papers have multiple authors, the author-wise features usually use the statistics indicators of a paper's authors to profile the paper, including mean, maximum, minimum, summation, median and variance of authors' attributes. \emph{Venues} refer to both journals and conferences where papers have been published. Although many indicators are designed for journal, we can also perform them on conferences in the same way. For example, the impact factor (IF) is defined as yearly average number of citations to recent articles published in that journal, which is only available for journals, especially journals indexed by \emph{Science Citation Index}. We can extend this definition to evaluate conferences in the same way. Besides, many network-based features are computed on the heterogeneous scholarly network constructed by author, paper and journal citation network and author collaboration network \citep{Shibata2012Link}.

Researchers try to portray the quality of papers from these three aspects. Some scholars take the structure of citation network and authors' cooperation network into account. Based on the citation network, Mcnamara et al. \cite{Mcnamara2013Predicting} put features of neighbours of the target paper into the prediction model. By adding some network structure features, e.g. the centrality and the PageRank score of nodes, Davletov et al. \cite{Davletov2014High} put forward a two-step method to predict paper citation. The method solved a classification problem at first. After determining the cluster of a given paper, this method computed the topological features of it and gave the prediction result by a polynomial regression. This method outperformed another model based on non-topological features as demonstrated in \citep{Yan2011Citation}.

This phenomenon reflects a problem with the family of methods based on extracting features. That is, there are many papers with very similar features but their citation accumulating processes are totally different. Although the likelihood of a paper being cited is mainly depended on its quality, many other factors will have influence on it. Maybe depicting the paper's quality is easy by the features of its author's reputation, but other influential factors are much harder to quantify. For example, Bornmann et al. \cite{Bornmann2013How} found that using the length of a paper and the numbers of paper's references as features can improve the prediction accuracy, but we cannot evaluate a paper's quality by such properties. Thus, finding useful features to improve the prediction remains an active research challenge.

The analysis above demonstrate the limitation of the prediction methods based on machine learning. Many efforts have been made to overcome these shortcomings. For example, many researchers study the correlation between the existing features and prediction target and try to find more relative features or feature combinations. Singh et al. \cite{Singh2017Understanding} analysed the early citers' impact on the target paper's citation accumulating. Statistical results showed that influential early citers (the authors of the target paper's citation within 1-2 years after publication) may have negative effect on the citation accumulating of the cited paper, which is also known as \emph{attention stealing}. Based on this, they predicted paper's citation count by using the influence of paper's early citers, together with the frequently-used paper-wise, author-wise and venue-wise features, and a group of features of citation context, which is also discussed in \cite{Hassan2017Identifying}.

\subsubsection{Other Approaches}
Paper citation prediction faces two problems: (1) paper's citation counts are unevenly distributed, with only a few papers are frequently cited; (2) if the history citation of a paper is unknown, the prediction accuracy will be limited. In this section, we will discuss the methods addressing these problems.

Chakraborty et al. \cite{Chakraborty2014Towards} treated the prediction task as a two-step problem, including a classification step using SVM and a regression step using SVR, to estimate a range for a paper's citation counts in the first step and then predict the exact citation count in that range. In the first step, all papers were divided into 6 different classes by its accumulative citation counts in the first five years since publication. Then a classification model was trained using paper-wise, venue-wise and author-wise features. The second step used the same features to train a SVR model to predict every paper's citation count. This method has achieved a satisfactory result because at the first step papers with similar citation counts are classified into the same set before predicting their citation, which can decrease the prediction error brought by the nonuniform distribution of samples. However it still suffers from the error in the classification step. According to their experiments, the overall classification accuracy is 0.78, which is relatively low. A similar method is proposed by \cite{Davletov2014High}.

Bhat et al. \cite{Bhat2016Citation} proposed a method to predict the classifications of paper citation without using the early citation as feature. They divided all papers into three classes according to citation: (1)papers without any citations (low-cited papers); (2) papers with citations between 1 and 12 (mid-cited papers); and (3) papers with citations more than 12 (high-cited papers). Then, they predicted the class of a target paper. The experiments showed that the accuracy is 0.87 when papers are divided into two classes and 0.66 when papers are divided into three classes. Because the features used do not include the citation information of target papers, this method can be used to classify a new published paper, despite the relatively low accuracy. Another innovation of this paper is that it used the Shannon entropy and Jensen-Shannon divergence to quantify the interdisciplinarity of authors.

\subsection{Time Series Models}
The accumulation of paper's citation can be modelled based on the hypothesis that a paper attracts each citation with a certain probability \citep{bai2017overview}. If the probability of paper being cited can be quantified, the process of paper citation accumulation can be described by a statistical model. This class of methods mainly focuses on modeling the distribution of a paper's citation counts accumulating process. Based on the knowledge of the distribution of paper's citation counts, researchers use time series models to fit real citation sequence of paper, including the log-normal distribution, the Poisson Process model, the Hawks Process model.

\subsubsection{Paper Citation Dynamics}
Radicchi et al. \cite{Radicchi2008Universality} claimed that the distribution pattern of paper citation is universal for all scientific fields subject to a scaling factor. However, in a later study, Waltman et al. \cite{Waltman2012Universality} concluded that different citation distribution patterns were observed for some fields with low average number of citations. In practice, most studies focus on only one research field to avoid this problem.

\begin{figure}[h]
  \centering
  \includegraphics[width=0.8\linewidth]{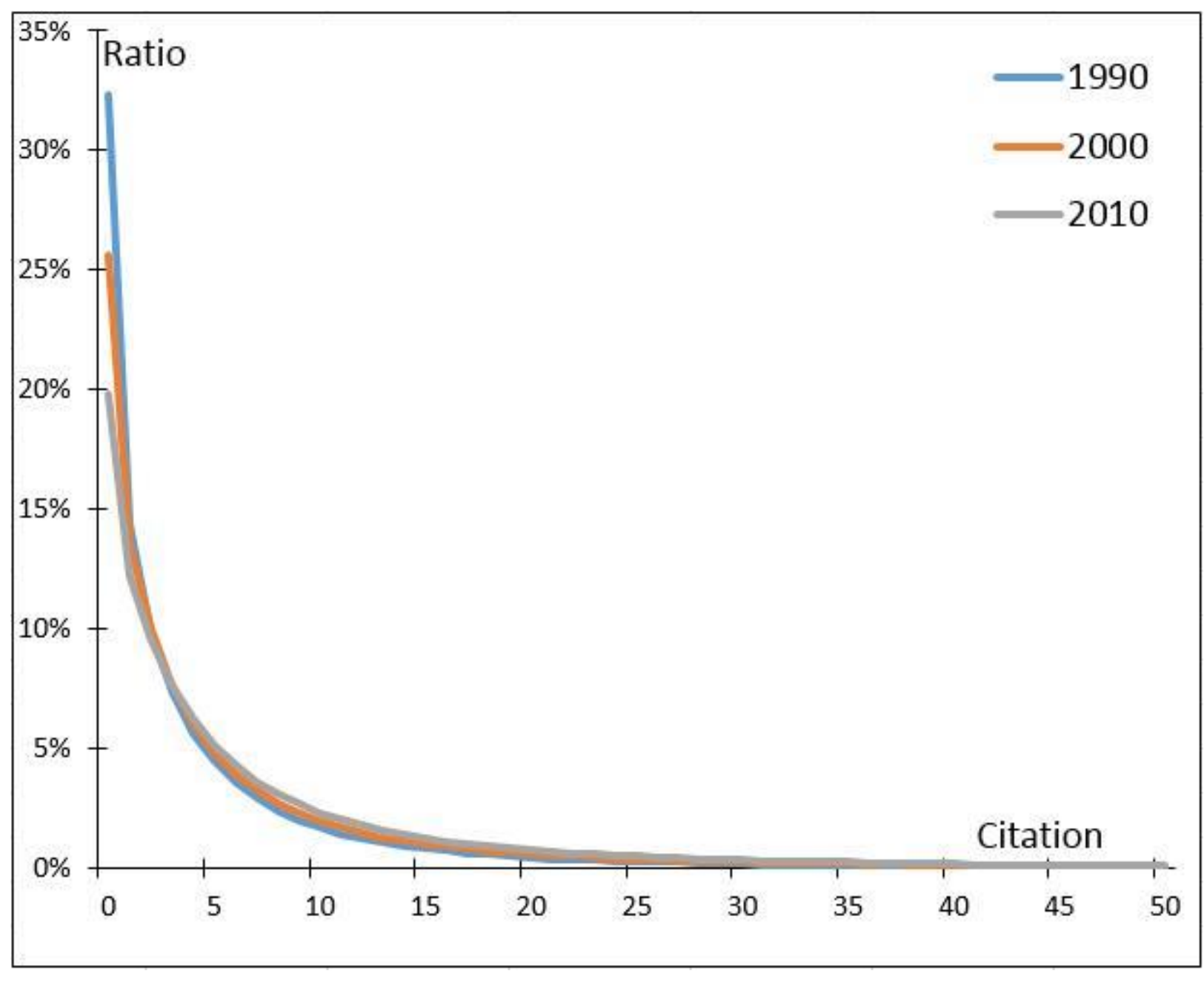}
  \caption{\label{Figure1} The citation distribution of papers in year 1990, 2000 and 2010 in APS dataset. The proportion of papers with certain citation counts is very close among 1990, 2000 and 2010. The proportion of papers that have been cited more than 50 times is very low, which is $1.1\%$ in 1990, $1.9\%$ in 2000 and $2.9\%$ in 2010. Thus, these parts of distribution are not shown in the figure.}
  \label{citation}
\end{figure}

Redner et al. \cite{Redner2005Citation} performed a statistical analysis on the APS family of journals from 1893 to 2004. The result showed that the accumulative paper citation follows a log-normal distribution. Following Redner's study, Emo et al. \cite{eom2011characterizing} also analyzed the citation distribution of APS papers.
They found that shifted power low distribution can fit paper citaiton better than log-normal.  %???
Their model considered a variant of linear preferential attachment that every paper had an exponentially decaying with time attractiveness that can be modeled by a heterogeneous power-law distribution. Wallace et al. \cite{Wallace2009Modeling} studied the citation distribution over 100 years using the data of Web of Scholar and came up with a conclusion that it followed a Tsallis distribution. Actually, the curves of citation distribution in both of their analyses are very similar but the log-normal distribution has a simpler form. Thus the citation distribution is always seen as log-normal. The citation distribution of the papers in APS dataset is shown in Fig.\ref{citation}.

\subsubsection{Modeling Paper Citation}
Preferential attachment theory is always used to model paper citation accumulation process.  %???
Wu et al. \cite{wu2014generalized} proposed a generalized preferential attachment model that considered the age of papers. Wang et al. \cite{Wang2013Quantifying} proposed WSB based on the knowledge that paper's citation accumulation follows a log-normal distribution. The method used three parameters, $\lambda$, $\mu$ and $\sigma$, to model the paper's citation and used the least square fit method to obtain their values. In this paper, the authors also gave a way to estimate journal's impact factor and the upper bound of paper citation count by their model. Although WSB works well on most data, it also has some defects. The experiments showed that the WSB model achieved a higher accuracy for low citation papers but the accuracy decreased for highly cited papers.

Wang et al. \cite{Wang2014Science} argued that the WSB model suffered from over-fitting. They had observed large mean prediction errors when using the WSB model to predict future citations of individual papers and made three main claims that, first, the value of parameters $\mu$ and $\lambda$ usually turned out to be very large; second, the mean absolute percent error (MAPE) of the result was always very large; and third, this method's performance can be worse than some naive approaches. To address these limitations, Shen et al. \cite{Shen2014Modeling} used Bayesian approach to obtain the values of parameters to avoid over-fitting.

Also in \cite{Shen2014Modeling}, authors proposed a Reinforced Poisson Process (RPP) model, which is an extension of the WSB model. In this paper, citation accumulating process is assumed to follow a Poisson distribution with a certain intensity. Based on their prior work, the function of intensity is log-normal, which is the same as \cite{Wang2013Quantifying}. Instead of optimizing the distribution function directly, they used RPP with prior to model the paper's citation sequence. Then, they performed a Bayesian prediction on the parameters to avoid over-fitting. Experiments showed that the RPP model outperformed the WSB model.

While using the Bayesian Inference model can avoid the over-fitting problem, the models above have an obvious limitation: they cannot account for exogenous ``second acts''. Most papers' annual citation would decline after a certain number of years. A ``second acts'' is that the annual citation surges after a long low-citation period. This problem can be handled by the preferential attachment theory by the way discussed in \cite{eom2011characterizing}. Xiao et al. \cite{Xiao2016On} proposed a model based on the Poisson process to try to overcome this limitation. Empirically, the accumulation of paper citation is mainly influenced by three factors. A paper has its intrinsic popularity which is decided by its publishing information, such as the authors' impact and the journal's impact. Majority of papers will gradually lose its popularity and impact over time, which is the same as the attractiveness decaying process in \cite{eom2011characterizing}. The proposed model also takes the Matthew Effect into account by adding a \textit{recency-sensitive} citation trigger. This Poisson process model obtained a better result when dealing with the ``second act''. Otherwise, the same method performed very well on predicting the citation of patents, as shown in \cite{Liu2017On}. Bai et al. \cite{bai2019predicting} introduced a new parameter to the preferential attachment to quantify the influence of early citers of a paper, which made the model capable of describing more complex citation accumulation process.

Another different point of modeling paper citation is the reference-citation duality. Golosovsky et al. \cite{Golosovsky2017Growing} made use of the duality and gave a predictive model in their recent work. Because a paper's references have been determined before the paper published, it is easier to analyse the references than citations. Via the duality between mean annual numbers of references and citations, researchers performed an analysis of the relationship of reference and citation's distribution. The authors used a mean-field method to get the approximate mean citation dynamics. To describe the individual paper's citation dynamics, the citations are divided into two classes, direct citation and indirect citation. The model proposed is a combination of these two parts of citation's probability distribution modeled by the Hawks process. The highlights in this paper are the analyses of reference-citation duality and the direct/indirect citation.

In addition to the models based on preferential attachment theory introduced above, many other old-school dynamics modeling methods are used in modeling and predicting paper citation. Min et al. \cite{min2018innovation} used a Bass diffusion model to describe the citation dynamics of papers published by 629 Nobel Prize winners. They used two parameters $p$ and $q$ to estimate paper¡¯s present status and the trend of citation accumulation. Poncela-Casasnovas et al. \cite{poncela-casasnovas2019large} used a randomization null model to discover the differences between citation patterns of high and low-impact papers. Kim et al. \cite{kim2017modeling} modeled the paper's citation accumulation with a Poison Process model that considering affinity between papers.

Although there are many studies for discovering the definite distribution of citation, it is not necessary to give a specific function. Cao et al. \cite{Cao2016A} made a hypothesis that if papers in the same field get the same numbers of citation in given years, they will have a similar citation distribution in the future. Based on this hypothesis, they proposed a simple method to predict paper's long-term citation. Given a paper's first three years' citation, they firstly identified several papers with the same citation sequence in the same journal or conference, and then uses the average of these papers citation counts in the following years to make a prediction. This method works like classical k-Nearest Neighbour classification, which doesn't require complex computation. The experiments showed that method proposed in \cite{Cao2016A} outperformed the WSB model.

\subsection{Link Prediction and the Application in Citation Prediction}
With the development of complex network theory, the study of link prediction also attracts many attentions. In the science of science, link prediction is always performed on author cooperation network, which will be discussed in the following sections. As a kind of complex network, paper's citation network also has been studied from the view of link prediction. The citation network is a directed network with strong temporality. Some researchers try to discover the evolution rules of the citation network to improve the link prediction accuracy, which is also helpful for understanding the mechanism of paper's citation accumulation.

\subsubsection{Link Prediction}
\begin{figure}[htbp]
  \begin{center}
    \includegraphics[width=1.0\linewidth]{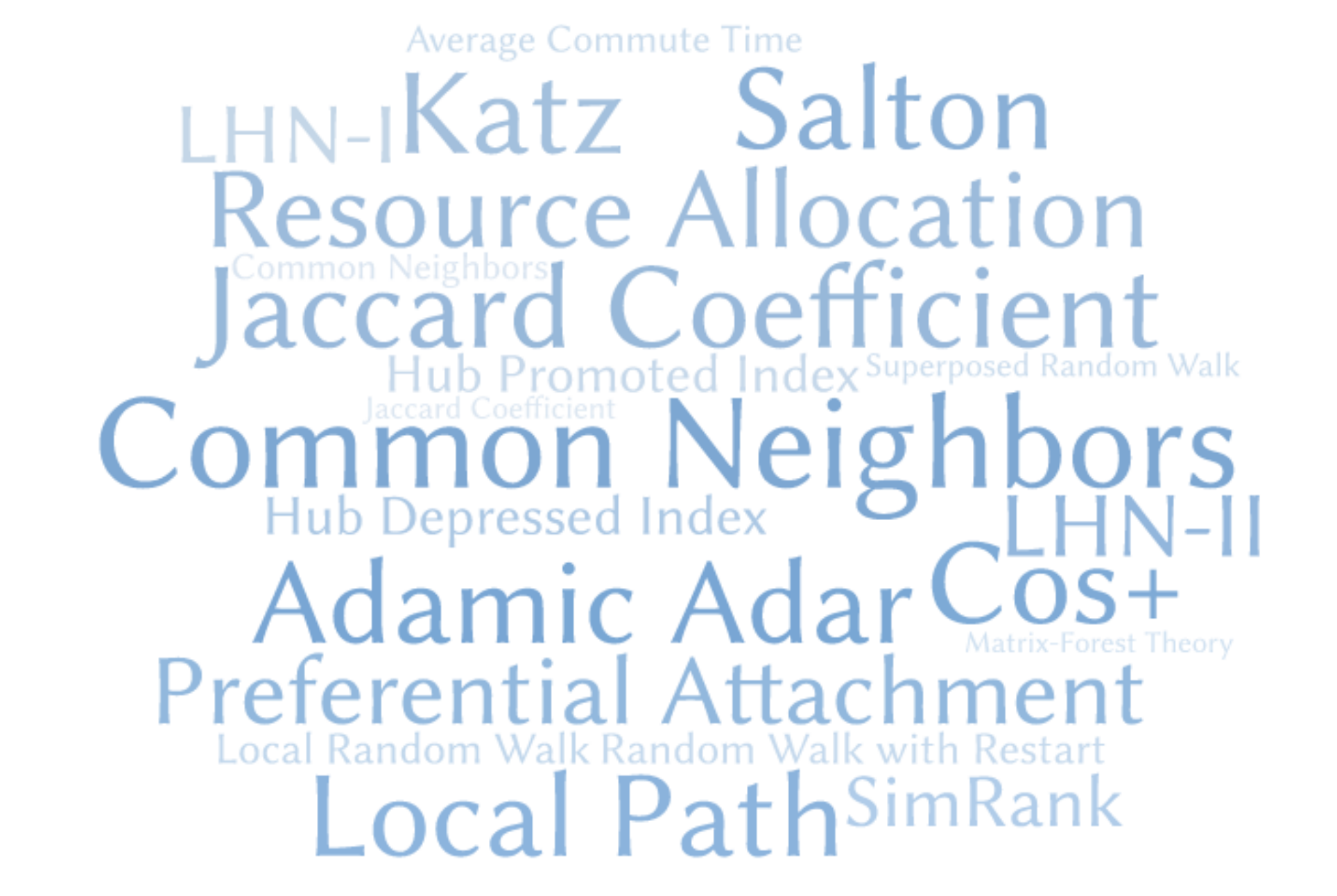}
    \caption{Frequently used indices for node similarity.}
    \label{Link}
  \end{center}
\end{figure}

Generally, link prediction method could be divided into two categories. One is based on nodes' similarity, and the other is based on maximum likelihood or probabilistic models \citep{L2011Link,Mart2016A}. The former is to calculate proximity measures from various dimensions. Fig.\ref{Link} shows the frequently used index evaluating nodes' similarity, including common neighbours (CN) \citep{b12}, Adamic-Adar index (AA) \citep{Adamic2003Friends}, Resource Allocation index (RA) \citep{Zhou2009Predicting}. They are computed for node pairs that are not connected at present in the network. Then, the proximity scores of all node pairs are used for performing the link prediction by an unsupervised method or a supervised method. The supervised method treats the prediction problem as a binary classification task, and the unsupervised method is to sort links in descending order by their proximity score. The probabilistic method for link prediction is to find and model the mechanism of network evolution, which is realized by hypotheses of the organizing principles of the network structure, e.g. hierarchical structure model and stochastic block model, or giving a statistical model to fit the evolution rule.

In paper citation prediction, the methods based on nodes' similarity are more popular. The prediction target is usually the existence of citation relationship between a given pair of papers, rather than the citation count. There are also methods that using the proximity measures as features to predict citation count by the machine learning models. These methods are similar to the prediction methods mentioned in the previous section.

\subsubsection{Link Prediction Used for Citation Prediction}
Link prediction is always used to predict the citation relationship between papers. Shibata et al. \cite{Shibata2012Link} used a supervised method to predict paper's citation relationship. The features used in their experiment included topological features, semantic features and paper's attribute features. In the experiment, Support Vector Classifier (SVC) was used as the classifier, and some existing citation pairs were extracted as positive instances. The negative instances were created randomly. The authors argued that given the entire citation network, their model can predict the existence of a citation with the probability ranged from 0.74 to 0.82. Jawed et al. \cite{Jawed2015Time} introduced a time-frame with the respect of topological features to predict links in directed network, using an unsupervised method. The model takes the evolution process of network with time into account, which is a characteristic of citation networks.

L\"{u} et al. \cite{Linyuan2016The} proposed the H-index of network nodes and analysed its relation to degree and coreness. Based on this work, Jia et al. \cite{Jia2017Improve} used the node's H-index replacing degree to compute some degree-based proximity measures. By using these H-index-based features, they performed a supervised method to predict the existence of a given citation pair, which yielded a promising result. Different with methods proposed above, Pobiedina et al. \cite{Pobiedina2016Citation} used the proximity measures as features to predict the paper's citation count directly, like the methods based on statistical learning. Besides the common features, the authors have defined a group of graph evolution rules (GER) and calculate the GERscore of nodes as a specific feature. The model classified the paper's according to their citation count at first, and then performed a regression task to predict the target paper's citation count.

\section{Scholar Impact Prediction}
\label{sec:3}
Scholar impact prediction is another significant problem in the science of science, and scholar-wise features such as the authors' impact correlates to the impact of a paper \cite{Singh2017Understanding}. Thus, there's a high degree of relevance between scholar impact and paper impact. Scholar impact prediction utilizes the same datasets (such as MAG) and shares similar prediction frameworks with paper impact prediction.

Predicting the impact of scholars can provide guidance to the scientific community on many aspects, such as the allocation of funds and the promotion of university faculty members.
In this section, we introduce indices used to evaluate scholar's impact and methods of predicting the future impact.

\subsection{Quantifying Scientific Impact}
\begin{figure}[htbp]
\begin{center}
\includegraphics[width=1.0\linewidth]{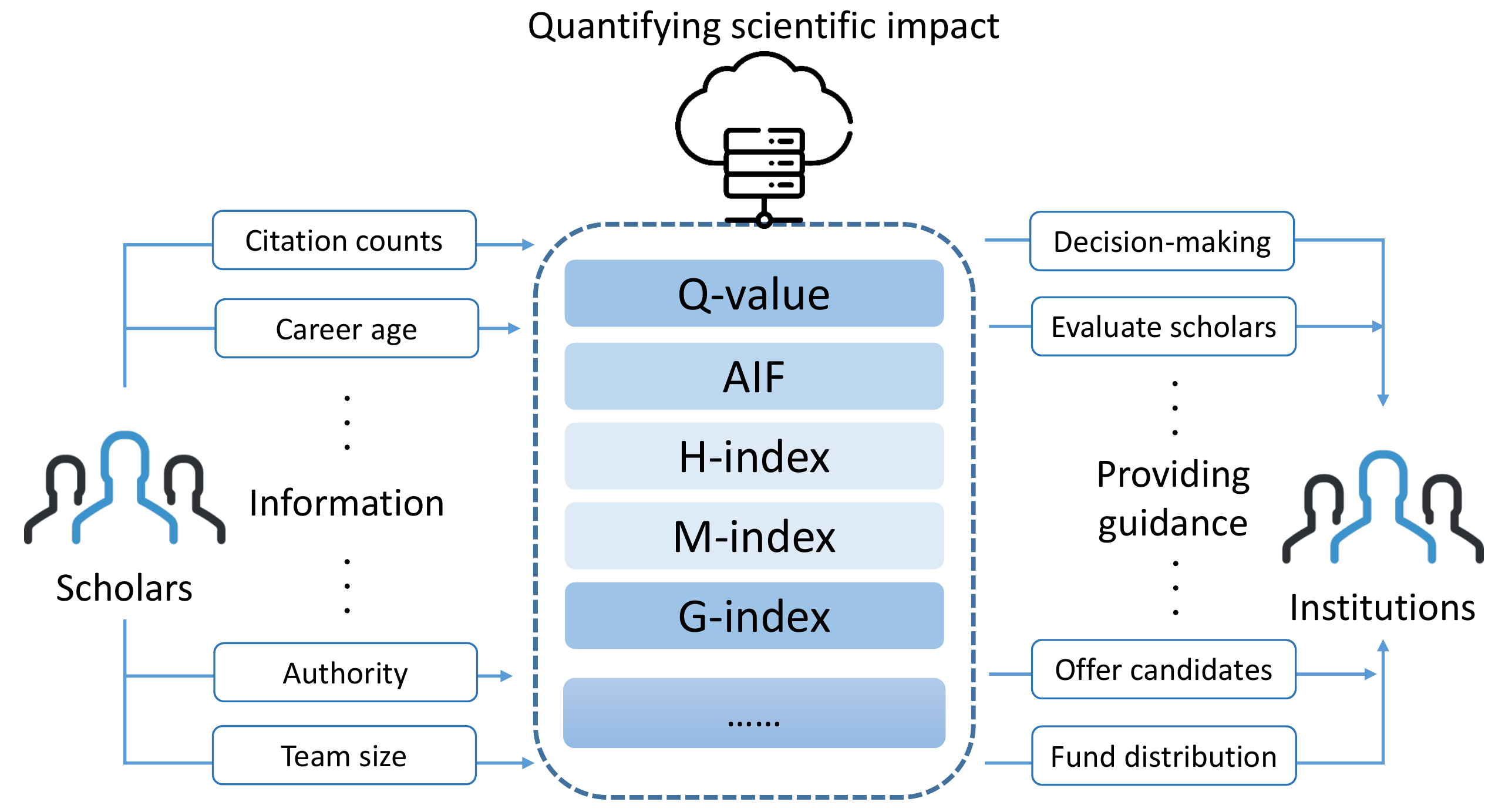}
\caption{The process of quantifying scientific impact.}
\label{quantify}
\end{center}

\end{figure}

Different from the impact of papers, scholar's impact is more difficult to evaluate. There are many indices for evaluating scholar's impact and the scholar impact prediction is also usually based on these indices, such as author impact factor (AIF) \citep{Pan2013Author}, Q-value \citep{a10} and H-index \citep{hirsch2005index}. In this section, we will introduce some commonly used indices for evaluating scholar's impact.
The process is described in Fig.\ref{quantify}.

AIF use the concept of journal's Impact Factor for assessing the author impact. AIF is defined as Function. \ref{equ:aif}, where $\Delta t$ is the time window of computing author impact factor, which is always set as recent 2 or 5 years. % ???
$PaperNum_y$ indicates the total number of papers published in year $y$ during the period $\Delta t$ and $Citation_i$ indicates citation counts accumulated by one of these papers $i$. It can capture trends and variations of the impact of the scientific output of scholars over time \citep{Pan2013Author}. However, because of AIF may change with a delay, it doesn't capture both of actual evolution of scholars' career impact and their productivity.
\begin{equation}
    AIF_{\Delta t} = \frac{\sum_{i} Citation_i}{\sum_{y \in \Delta t} PaperNum_y}
    \label{equ:aif}
\end{equation}

In order to find a proper indicator to quantify scholar impact, various studies have been conducted.
Sinatra et al. \cite{a10} and Nezhadbiglari et al. \cite{a11} found that the most influential papers were randomly generated in author's career by analysing the publications of scientists in various fields.
Therefore, Sinatra et al. \cite{a10} proposed a quantitative model (Q model) to study the role of scientists in scientific research based on random impact model (R model).
R model assumes that influence of each published paper is generated from the same distribution by random. Thus, the difference between various scientists is their whole productivity throughout their scientific career. % ???
To compensate this shortcoming of the R model, authors assumed that there is a unique parameter $Q_i$ for each scientists which quantifies the ability of them to improve a project selected randomly with potential $p_a$ and publish a paper of impact $c_{ia} = Q_i \cdot p_a$. Therefor, the value of $Q_i$ can also be used as a factor to quantify impact of scientists, called Q-value.
% each scientist randomly selects a project with potential pa and improves on it with a factor Qi that is unique to the scientist, resulting in a paper of impact The Q model analyses each scientist's personal impact by disordering papers' publication time sequence. %???
According to the model definition, the Q-value can be calculated by using publication sequence, which is defined as Function. \ref{equ:q}, where $c_{10, i}$ is the average citation of papers published by scholar $i$ in recent 10 years and $\mu _p$ is the average potential impact of these papers.
% Q-value might be influenced by various factors, such as a consistent high performance throughout the scholar's career, instead of a single paper for example. Therefore,
% Q-value can predict the career impact to some extent and Q model is a good choice in the aspect of quantifying scholar impact.
\begin{equation}
    Q_i = e^{\langle \log c_{10, i}\rangle - \mu_p}
    \label{equ:q}
\end{equation}

However, using the features summarized from quantifying scholar impact to predict scholars' scientific impact needs to be further tested.
Mazloumian \cite{a2} found that the annual citations is a better predictor at the time of prediction by testing 10 citation indicators, such as H-index, M-index and G-index.
Later, Havemann et al. \cite{a7} found that the output index of papers was of little significance to predict rising stars by testing bibliometric indicators.
While a large number of papers are written by multiple co-authors, it is difficult to quantify individual contribution hence the associated impact.
Therefore, P{\~o}der \cite{a15} assumed that a single author can be measured and predicted by the patterns of publication and citation processes, which are modelled separately. In \cite{a15}, the authors also assumed that each scholar's productivity was the same. If more papers were published, the contribution of those authors to the entire group will be clearer because important authors could appear more frequently. So it will be more accurate to quantify each scholar's productivity.
Thus, there's a need to identify feasible criteria to better quantify and predict the future impact of scholars, rather than simply using common indicators for measurement and evaluation.

\subsection{Predicting Scholar Impact}

We have discussed some indicators to quantify scholar impact. Among these indicators, H-index is most widely used. Hirsch \cite{hirsch2005index} first introduced H-index and studied the predictability of it. He found that the H-index was powerful for predicting the future impact of scholars \citep{Hirsch2007Does}. H-index quantifies productivity and quality of scholar's research outputs, and subsequently many other indicators are proposed based on this framework \citep{Pan2013Author}.

Many researchers try to predict the future impact of scientists by predicting their future H-index. The prediction can be performed directly or indirectly.
The direct way uses regression models to predict scholar's future H-index \citep{a3}. Mistele et al. \citep{mistele2019predicting} proposed a neural network approach to predict scholar's citation and H-index. The indirect way is to predict scholar's future H-index by quantifying the quality of their published papers \citep{a13}.

\begin{figure}[htbp]
  \begin{center}
    \includegraphics[width=1.0\linewidth]{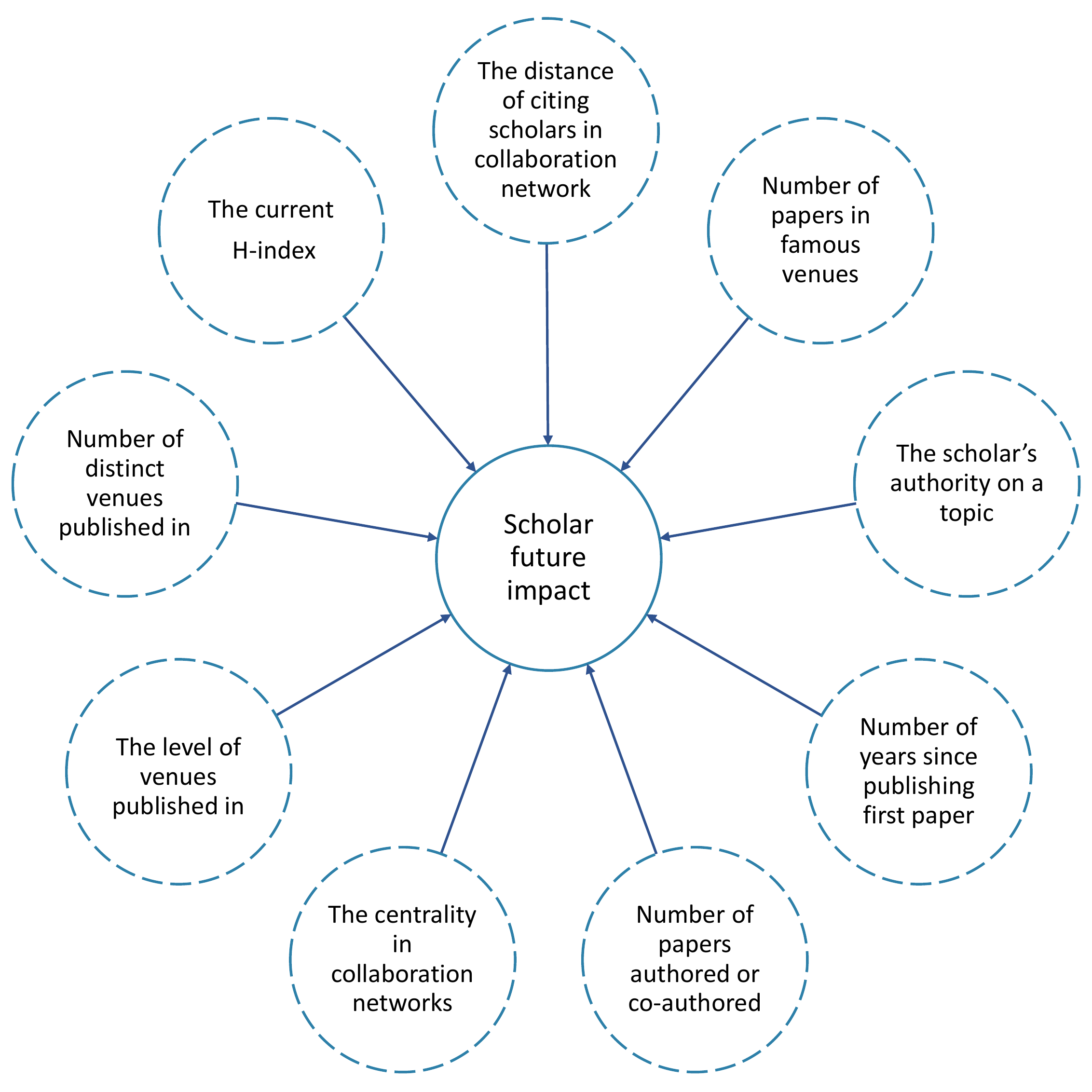}
    \caption{Features used to predict scholar impact \citep{a3,a6,a13,a15,Bras2011A,sarigol2014predicting}.}
    \label{factors}
  \end{center}
\end{figure}

Acuna et al. \cite{a3} predicted scholar's impact by considering the following features of the target scholar: the current H-index of the scholar, the number of years since the first paper being published by the scholar, the number of different journals that the scholar has published papers before, and the number of papers published in prestigious journals in the target scholar's research field.
Their studies include five representative journals for neuroscientists, including \textit{Science}, \textit{Nature}, \textit{Proceedings of the National Academy of Sciences}, \textit{Nature Neuroscience} and \textit{Neuron}.
However, their method had some drawbacks. It does not manage well the uncertainty surrounding careers at different levels and stages, such as career age constraints, as scientists tend to have more career volatility in early stage, leading to inaccurate prediction of young researchers \citep{a6,a13}.
In order to solve this problem, Dong et al. \cite{a13} divided prediction into two tasks. One is to use the current scientific impact of scholars to predict H-index after a predetermined period of time, and the other is to add a new paper to see how their future H-index changed.
Their method predicts the H-index of scholars by constantly joining the paper. Therefore, some factors that influence the author's future impact are extracted from the H-index and citation counts \citep{ a2,a12,a14}, which can help the prediction of scholar impact.

Many relevant factors have been explored for predicting the future H-index of scholars. Topical authority and publication journals (conferences) are important for whether the paper will have a high impact on the future H-index of its authors \citep{a13,a17}, instead of the popularity of the topics and the impact of co-authors. Some common features which contribute to the prediction of scholar impact are shown in Fig.\ref{factors}.

Many models are used to predict scholar's impact from various aspects \citep{gogoglou2016scientist, poder2017framework, ayaz2018predicting}. In various studies, it has been found that using non-linear models to predict the future scientific impact of scholars can obtain good results \citep{a14}.
Dong et al. \cite{a13} demonstrated the predictability of scientific impact through prediction results of some classification models including SVM, LR, RF and radial basis function network (RBF).
They found that the prediction in a short period of time would be more accurate.
However, Weihs et al. \cite{a14} found that the prediction of the next ten years is also feasible.

Like papers, the evolution rules of scholar's impact in different research fields is different. Identifying the difference between research fields helps improving scholar impact prediction.
For the diversity of the fields, the use of knowledge flow maps (KFMs) of science will be more convenient. The existing maps of science include physics, computers and sociology \citep{a19}. The commonly used methods based on KFMs include co-citation, direct citations, and bibliographic coupling. The career trajectories are also important parts of KFMs \citep{a20}. Guevara et al. \cite{a12} used career trajectories to predict scholar impact. The first step is to categorize publications by the fields, and then maps papers into their associated classes. The mapping step is based on bibliographic coupling in order to predict the forthcoming disaggregation in a research field.

The traditional citation count can be used to measure the popularity of a scholar in their career, which is widely used to measure the popularity of scientific publications \citep{a21,a22,a23,a24}. Citation counts of scholars are defined as the total citation accumulated by all their publications. Using the number of citations to quantify the research contribution of scholars is accurate to some extent \cite{a2,a21}. Nezhadbiglari et al. \cite{a11} proposed a model to predict scholars' popularity through predicting their citation counts. This model has two parts. The first part extracts the current popularity trend for each scholar, and then predict their popularity trends. They extracted the trend of influencing authors' popularity through K-SC clustering algorithm. This algorithm keeps updating the centroids until it converges, and the time series of each cluster follows this trend. This prediction is based on the identified cluster centroids to calculate the academic characteristics of this period. They adapted models on predicting user generated content (UGC) popularity to predict the trend of each scholar, using different scholar features as the input to train the classification method \citep{Bras2011A}.

\subsection{Identifying Rising Stars}
In this section, we will introduce the methods about identifying rising stars proposed in recent years. Identifying the academic rising stars is the other part of predicting scholar impact. It is pivotal for the science community to provide a strong guidance to the new generation of academic leaders.
If there is a rapid change in a scholar's impact under a rising trend, we define such scholar as a rising star. To detect the academic rising stars is crucial for predicting the impact of scholars.

Finding rising stars in co-authorship network is an emerging research area \citep{b26,b27,b28,b29}, and the idea about this approach was first formulated by PubRank \citep{b27} as shown in \ref{eq:pr1}.
\begin{equation}
  PubRank(A_{i}) = \frac{1 - d}{n} + d \cdot X,
  \label{eq:pr1}
\end{equation}
and
\begin{equation}
  X = \sum_{j = 1}^{\left | v \right |}\frac{W(A_{i}, A_{j}) \cdot \lambda (A_{i}) \cdot PubRank(A_{j})}{\sum_{k = 1}^{\left | v \right |}W(A_{k}, A_{j}) \cdot \lambda (A_{k})},
  \label{eq:pr2}
\end{equation}
where $W$ is the influence of author $A_i$ on author $A_j$, which is defined as the proportion of numbers of co-authored papers to papers authored by $A_i$, $\lambda$ is the publication quality score for an author, $n$ is the total number of authors, $\lambda$ is static ranking information of publication quality \citep{a18}, and $d$ is damping factor and its value usually defined between 0 and 1.
However, PubRank only mentioned static ranking of publication, authors and papers mutual influence.

Later, StarRank enhanced PubRank in two aspects, considering co-authors' mutual influence, and using dynamic ranking lists of publications $\lambda (dpq)$ instead of static ranking \citep{b26}. It also considered the first author as the major contributor, the $AOWI$ is author order weight based mutual influence, which is defined as the proportion of authors' co-authored contributions to the total contribution of author $A_j$. The function of StarRank is as follows:
\begin{equation}
  \mathit{StarRank}(A_{i}) = \frac{1 - d}{n} + d \cdot S,
  \label{eq:sr1}
\end{equation}
and
\begin{equation}
  S = \sum_{j = 1}^{|v|}\frac{AOWI(A_{i},A_{j}) \cdot \lambda (dpq) \cdot \mathit{StarRank}(A_{j})}{\sum_{k = 1}^{|v|}AOWI(A_{k},A_{j})\cdot \lambda (dpq)},
  \label{eq:sr2}
\end{equation}
where $\lambda (dpq_{i}) = \frac{1}{\left | p \right |} \cdot \sum_{i = 1}^{\left | p \right |}\frac{1}{\alpha ^{Entropy of Journal}}$.

\begin{figure}[htbp]
  \begin{center}
    \includegraphics[width=1.0\linewidth]{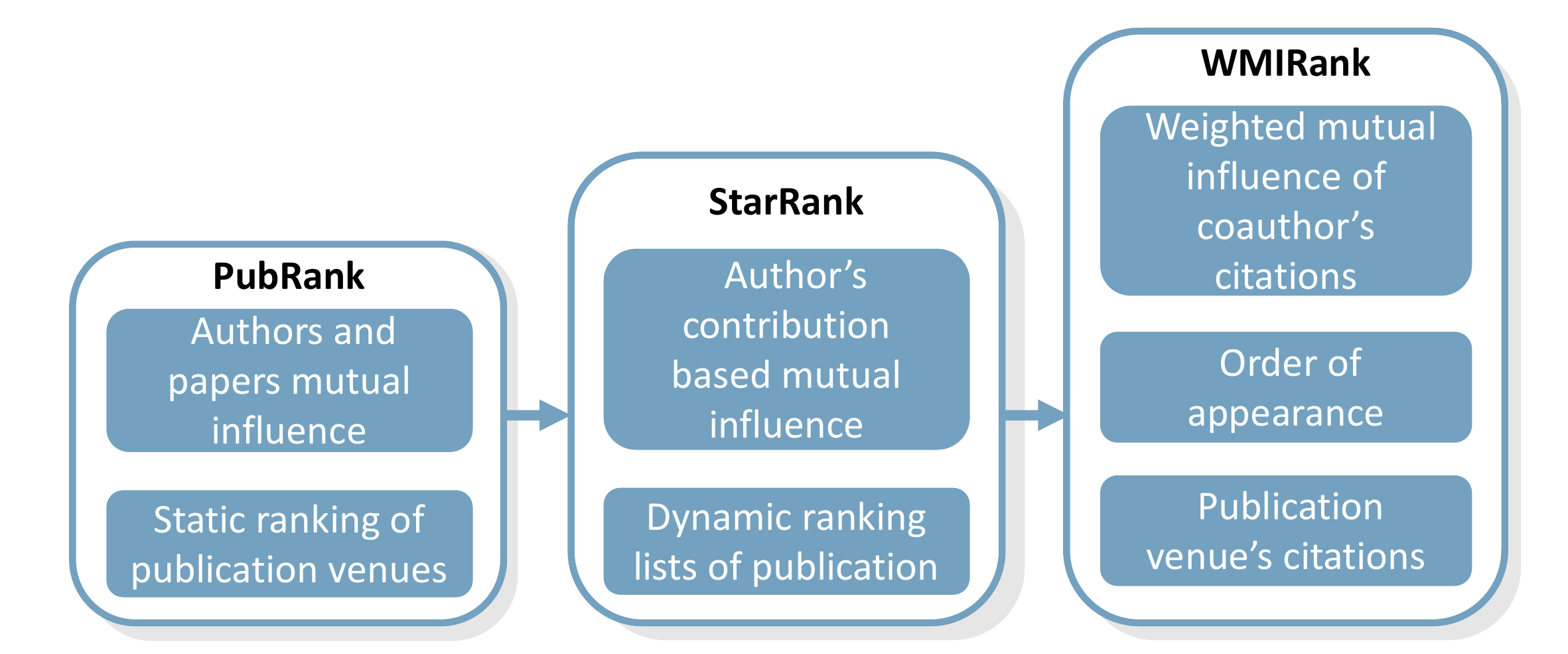}
    \caption{Overview of the evolution about how to find rising stars in co-authorship network \citep{b9,b27,b26}.}
    \label{WMIRank}
  \end{center}
\end{figure}

In some cases, rising stars could be considered as the top scholars in the future scholar rank system. In order to rank scholars more accurately, Daud et al. \cite{b9} proposed Weighted Mutual Influence Rank (WMIRank), defined as
\begin{equation}
\mathit{WMIRank}(A_{i}) = \frac{1 - d}{n} + d \cdot T \cdot \mathit{WMIRank}(A_{j}),
\label{eq:wr1}
\end{equation}
and
\begin{equation}
T = \sum_{j = 1}^{\left | v \right |}\frac{WI(A_i, A_j)}{\sum_{k = 1}^{\left | v \right |}WI(A_k, A_j)},
\end{equation}
where $WI(A_i, A_j) = CACWI(A_{i}, A_{j}) \cdot CAOWI(A_{i}, A_{j}) \cdot CAVWI(A_{i}, A_{j})$. It is combined with citations based mutual influence $CACWI$, co-author's order based mutual influence $CAOWI$ and co-author journals' citations based mutual influence $CAVWI$. If a co-author has more citations, they will have more impact on their collaborators. Thus, WMIRank considers this regulation as the first step. Generally, the first author in a paper refers to the person who produces the highest contribution. Therefore, WMIRank applied first author order as the main idea for the computation of influence of an author on other authors. If a co-author has more papers published in journals which have high impact and higher citations, the author will have more influence. Hence, WMIRank score is calculated by using these three factors to rank each author. The evolution of finding rising stars is shown in Fig.\ref{WMIRank}.

According to \cite{Seglen1997Why}, although publication journals affects the paper's visibility, impact factors can not reflect the difference in citation count of papers published on it. Some scholars argued that citation count is a better indicator of the paper's impact \citep{b29,Seglen1997Why}. Therefore, Panagopoulos et al. \cite{a17} analysed the performance of scholars in quality and quantity of their publications, the sociability and the ability about cooperating with famous scholars. Instead of simply performing binary classification on the scholars, they proposed the \emph{decay factor}, a measure of their success based on the oldness of their publication and the oldness of the citations they received. Li et al. \cite{a9} proposed an algorithm called Impact Crystal Ball (iBall) that predicted the long-term scientific impact of early scientists' career. This algorithm can be summarized as regression and classification model. Ning et al. \cite{ning2017social} proposed a method that used the weights of a fully connected neural network trained by scholars' attributes in the co-authorship network as features, which they called it \emph{Social Gene}, to predict rising stars. Later, Ding et al. \cite{ding2018rising} found that replacing fully connected neural network with decision trees as the feature extractor can get a better performance.

\section{Author Collaboration Prediction}
\label{sec:4}
Team-based research is a popular practice nowadays that leads to papers with multiple co-authors. Effective cooperation is conducive to promoting both scholars' and their papers' impact and many author collaboration based features are used in predicting paper and author's impact \cite{hurley2013deconstructing, sarigol2014predicting}. Better understanding the evolution of scholar's collaboration is helpful for the study of the evolution of paper and scholar's impact.

Scholar's collaboration is always modeled by networks. With the development of BSD, researchers are able to analyze the patterns of scholar's collaboration network based on a large amount of data while considering the attributes of papers and scholars \cite{a5, kong2019academic}. Some problems about how collaboration network evolves and how to promote scholars' collaboration still need to be solved. In this section, we will introduce recent achievements in author collaboration prediction.

% With the development of BSD, the number of researches about co-authorship network have increased \cite{a5}.
% Effective cooperation is conducive to the common progress of scholars in co-authorship network which is formed by papers and authors. Some problems about how co-authorship network evolves and how to promote scholars' co-authorship still need to be solved. These problems can be solved by using link prediction based on the structure of co-authorship network.

\subsection{Methods Based on Link Prediction}
Predicting the possible cooperation among scholars can provide good advice for better research collaboration in the future \citep{b7}.
% In co-authorship network, we need to predict the relationships between various members in the near future.
This problem can be simplified to the link prediction problem based on the proximity of nodes in the co-authorship network \citep{b12}.

% Some studies suggest that trust is essential for successful cooperation in co-authorship network \citep{b21,b22,b2}. Therefore,
Ghasemian et al. \cite{b2} considered the average of Jaccard similarity, degree centrality, and scholar's rank in terms of extracting valid features.
The use of these characteristics of the link structure between scholars can obtain better results.
They aimed to relate communities and motivate the CSCW (Computer Supported Cooperative Work) to form research collaborations successfully by considering more beneficial features, because a success collaboration means that the collaborators try their best to accomplish their primary goals. Thus, the problem can be solved in terms of considering which features will determine the scientific collaboration success \citep{b2}. Therefore, they computed the relations among the scholars and the rank of scholars. These two features are used to represent the collaboration relations by using hypergraph, which can capture the relations among scholars and the subject categories of their papers in each collaboration. But in their work, some activities among the scholars are ignored, such as grants and published books. They only considered the papers which scholars have published. Medina-Acu\~{n}a et al. \cite{medina2019link} used five classic classifiers (SVM, LR, DT, Neural Network and k-NN) with several well-known relational features, including common co-authors, Jaccard¡¯s coefficient and Adamic-Adar index to predict future links in scholar cooperation network.

Link structures can also be used for scientific predictions of cross-domain research collaborations.
Guo et al. \cite{b8} proposed a Hybrid Graph Model by using explicit co-author relationships and implicit co-citation relationships. They adopted supervised methods instead of unsupervised methods, and combined various features with different coefficients by training data. They proposed a historical collaboration information based model and a mining co-citation information based model. Then they combined these two models and built a Hybrid Graph Model by grasping both explicit and implicit information.
In this work, they discovered cross-cutting problems can be solved by using a Hybrid Graph Model, and the information of citation is useful for scholarly recommendation.

There are some problems with the traditional link prediction, such as does not consider the weight of the links generated at different past times \citep{b7}, cannot work well on some specific network \citep{b23}, and cannot found the co-authorship evolution between groups.
We will introduce some new algorithms to solve these problems to improve link prediction.

\subsection{Topological and Semantic Measures Based Methods}
The network is dynamically evolving. As time goes by, new nodes will be added into the collaboration network. There are lots of approaches to settle this problem, on the grounds that using topological information is better than using a random predictor.

Huang et al. \cite{b7} proposed a method taking time-varied weight into account. In this model, the researchers are represented by nodes in a collaboration network.
The method based on time-varied weight is to redefine these four indicators, and considers the weight of time and attenuation factor. They compared the performance of these features with the method based on time-varied weight and found that the latter had a higher AUC,
which may be due to similarity indicators do not consider time and dimension.

Taskar et al. \cite{Taskar2003Link} defined a joint probabilistic model by using discriminatively trained relational Markov networks to predict the evolution of co-authorship network. Kashima et al. \cite{kashima2006parameterized} proposed a parameterized probabilistic model to predict network evolution. However, these probabilistic models used to solve the link prediction problem are expensive to compute and do not perform well in large networks \citep{b11}.
Evans et al. \cite{b11} suggested that scientists working in the same institution or in related institutions were more likely to collaborate in order to reduce cost.
They used the topological and semantic approach to analyse the evolution of social networks to predict new co-authorship.
They found that scientists preferred to collaborate with colleagues from their own institutions.
However, their analysis only applies to collaborations within the boundaries of a single country and discipline.

Another study found that the Katz metric is a good feature for predicting future links \citep{b12}.
Katz is based on graph paths instead of obtaining values through neighbourhood analysis.
On the grounds of the volume of exacted words in scientific papers and projects, it is significant to solve polysemy and synonymy problem when doing matching between the words \citep{steyvers2004probabilistic,rosen2010learning,yang2013multilevel}.
Therefore, Str{\"o}de et al. \cite{b6} predicted new relationships through a metric composed by two parts. The one is to analyse objects content involved in social interaction to look for individuals who have common interests. And the other is to examine the network topology to look for unbalanced social structures and then balance it.
The Semantic Metric can extract main terms of their publications while comparing with topological and semantic approach. These main terms can verify and consolidate the common interests of researchers.

\subsection{Clustering Based Link Methods}
Node-property-based and network-structure-based approaches can obtain good results for the recommended collaborators by link prediction. However, there are some problems have not been solved. One is that these methods are not accurately modelled with link prediction method, and the other is that they do not work well on some specific networks \citep{b22,b23}.

Ma et al. \cite{b4} found that using the clustering method to mine the grouping trend of nodes and using clustering index (CI) to predict the potential of co-authorship networks can obtain better results than traditional network metrics. This method can better reflect the features of the co-authorship network, and the result is more reliable.

In the same cluster, researchers who are closer to the edge of the field are more willing to work with researchers at the center to increase their influence. And researchers whose positions are closer to the edge of the field are more willing to work with central researchers in another cluster to remedy their lack of knowledge. Therefore, there is a tendency for researchers who are closer to the edge of the field to cooperate with central researchers. The adjacency matrix of the co-authorship network is taken as the input of the spectral clustering algorithm, and the first turning point over the curve of sum of variance is used as the number of clusters in the co-authorship network \citep{b4}.

\subsection{Co-authorship Based Methods}
A hybrid mechanism has impact on the co-authorship evolution \citep{b3,b8}, but further research is needed on how different groups promote co-authorship evolution. Some studies showed that common interests played a significant role in the co-authorship network \citep{b6}. L{\"u} et al. \cite{L2011Link} measured the common interests of co-authors by using the network-based approach, which can measure common interests through scholars' shared co-authors. Hasan et al. \cite{Hasan2006Link} also explored common interests by using the content-based method.
From these studies, it has been found that the co-authorship evolution is driven by a variety of mechanisms.

In heterogeneous networks, Zhang \cite{b3} found that the co-authorship evolution mechanisms were based on various relations with their measurements. Not all research collaborations lead to co-authorship, and not all co-authors share a common research process.
In order to improve meta-path based model, Zhang \cite{b3} proposed multi-relation based on link prediction in heterogeneous bibliographic networks. Each mechanism and their combinations are used as predictors. They found that the mechanism denoted by the predictors with weights is the most suitable mechanism.

\section{Open Issues and Challenges}
\label{sec:5}
In previous sections, we have discussed the three main tasks in the data-driven prediction in the science of science. Lots of methods and applications are investigated, including the methods to predict paper's citation count, author's impact, author's cooperation and some metrics used to evaluate the quantity of scholar's outputs or the impact of scholars. Nevertheless, many open issues in this field remain unsolved or controversial. In this section, we give a brief introduction to such issues and discuss potential directions for future research.

\subsection{Differences in Research Fields}
It is clear that many attributes are different in different research fields, e.g. the number of papers published per year, the number of active researchers, the average number of references \citep{Lima2013Aggregating,Kaur2013Universality}. Thus, the study in the science of science differs across research fields. Because the datasets used by researchers are often limited to a certain field, the citation or cooperation across fields get missed. How to normalize the difference of research fields involves understanding the development trend and rules of different fields, which is beyond a simple statistical problem that can be analysed using statistical techniques. Discovering the rules of citation or cooperation across fields also is helpful to know the rise of interdisciplinary topics, which is useful for researchers to know the tendency of science development in advance.

\subsection{Self-citation}
Since the citation count of papers is a basic metric to evaluate research output's quantity and scholar's impact, some scholars are trying to promote their impact by citing their own papers, which makes the evaluation system unfair \citep{Ioannidis2015A,bai2017role}. Although it may have no influence on the prediction result, the negative impact of scholars may get extended. While not all self-citations are undesirable, it is subject to debate how self-citations should be count towards impact, and feasible assessment mechanisms are yet to be sought.

\subsection{Discover the Evolution Pattern of Scholarly Networks}
Link prediction, which predicts the emergency of new links or discovers unknown links, usually focuses on the structural evolution of the networks \citep{Zhou2009Predicting,L2011Link}. By performing link prediction methods on the citation network \citep{Shibata2012Link, Jawed2015Time, Jia2017Improve} and co-author network \citep{b3,b4,b6}, researchers can know these network's structural attributes better, which is helpful to grasp the evolution rules of the scholarly networks. Knowing the evaluation rules can also make it easier to build a predictive model for predicting paper's citations.

\subsection{Prediction Using Altmetrics}
With the development of the social media, such as Twitter, Facebook, micro-blog, the scientific community is an open society. Nowadays, increasingly research results or papers are spreading on the social media, which is helpful to promote a scholar's impact. The times of downloads, sharing or commenting of papers on an online forum have already been used to evaluate the research outputs, which is known as altmetrics \citep{Piwowar2013Altmetrics}. There are some researches on the predictability of the altmetrics \citep{Brody2005Earlier,Eysenbach2011Can,Ringelhan2015I,Peoples2016Twitter}, but some researchers argued that the altmetrics only have low correlations with the paper's citation counts \citep{Thelwall2013Do,Shema2014Do}. The study using altmetrics to predict the academic impact still requires further exploration.

\subsection{Prediction Paper Impact Using Deep Learning}
Deep learning is a branch of machine learning with growing popularity in the recent past. It has found tremendous potential in image classification, natural language process, machine translation and speech recognition. However, in paper impact prediction, there are limited studies using deep learning. Yuan et al. \cite{yuan2018modeling} proposed a prediction model using recurrent neural network with long short-term memory units. Recently, Abrishami et al. \cite{abrishami2019predicting} proposed a deep learning approach using RNN and auto encoder to predict paper's long-term citation. It's worth trying more to apply deep learning to predicting the impact of papers.

\subsection{Assessment of Scholar Success}
There are lots of indicators to measure scholar impact, but these indicators also have various shortages. Therefore, we need a standard to restrict and measure the level of scholars' success. If we want to quantify and predict a scholar's impact, we should have a uniform standard. In the scientific community, there are many famous indicators, such as H-index, \emph{g}-index . Different indicators have different effects. So far H-index is always regarded as the \emph{de facto} indicator, yet it cannot evaluate scholars in every aspect. There is a need for developing a standard indicator for measuring scholar impact. It means that scholar's success also needs a fair and quantifiable standard.

\subsection{Methods for Patent Impact Prediction}
Patents are another kind of research output besides papers. Many scholars will apply for patents for the new methods they put forward in their papers. Similar to papers, there are also abundant citation relationships among patents. Compared to papers, patents have more practical and commercial value. Predicting the future success of technology can tell us which technologies will be increasingly utilized in some fields, or which technologies are suitable to invest. Nowadays, some academic databases, such as Microsoft Academic Graph, have also included information of patents, which brings convenience for research of patent impact prediction. There are some similarities between the citation pattern of patent and paper and some prediction methods for paper citation can also be applied on patents \cite{Liu2017On}. But there are still limit researches on patent impact prediction and finding methods about predicting patent impact is still a promising and challenging research topic.
% Limited investigation in predicting the success of patents has been observed. Patent's power and expansion potential are regarded as technology scope indicators \citep{c1}. Predicting the future success of technology can tell us which technologies will be increasingly utilized in some fields, or which technologies are suitable to invest. The study on predicting future success of technology by predicting patent impact is complex \citep{c1}. Thus, finding methods about predicting patent impact is a promising research topic.

\section{Conclusion}
\label{sec:6}
In recent years, data-driven studies in the science of science has attracted growing attention in the scientific community. Many scientists have realized the significance of paper impact, scholar impact and author collaboration prediction. Predicting paper citation counts can improve the assessment of academic achievements. Predicting scholar impact and finding rising stars can provide useful guidance. Studying the underlying mechanisms of collaboration evolution can provide a comprehensive understanding of co-authorship formation and evolution.

This survey conducts an in-depth study of data-driven prediction in the science of science from three main aspects.
Opportunities abound for scientists to understand the evolution of the science of science in new environment. % ???
% Moreover, this survey shall also foster exploration in the development of methods and applications about data driven prediction in the science of science.

%\section*{References}

\bibliography{mybibfile}

\end{document}